\documentclass[fleqn,10pt]{wlscirep}
\usepackage[utf8]{inputenc}
\usepackage[T1]{fontenc}
\usepackage{amsmath}
\usepackage{braket}
\title{Improving solutions by embedding larger subproblems in a D-Wave quantum annealer}

\author[1,2,*]{Shuntaro Okada}
\author[2,3]{Masayuki Ohzeki}
\author[1]{Masayoshi Terabe}
\author[1]{Shinichiro Taguchi}
\affil[1]{Electronics R \& I Division, DENSO CORPORATION, Tokyo 103-6015, Japan}
\affil[2]{Graduate School of Information Sciences, Tohoku University, Sendai 980-8579, Japan}
\affil[3]{Institute of Innovative Research, Tokyo Institute of Technology, Yokohama 226-8503, Japan}

\affil[*]{shuntaro\_okada@denso.co.jp}

\begin{abstract}
Quantum annealing is a heuristic algorithm that solves combinatorial optimization problems, and D-Wave Systems Inc. has developed hardware implementation of this algorithm.
However, in general, we cannot embed all the logical variables of a large-scale problem, since the number of available qubits is limited.
In order to handle a large problem, {\tt qbsolv} has been proposed as a method for partitioning the original large problem into subproblems that are embeddable in the D-Wave quantum annealer, and it then iteratively optimizes the subproblems using the quantum annealer.
Multiple logical variables in the subproblem are simultaneously updated in this iterative solver, and using this approach we expect to obtain better solutions than can be obtained by conventional local search algorithms.
Although embedding of large subproblems is essential for improving the accuracy of solutions in this scheme, the size of the subproblems are small in {\tt qbsolv} since the subproblems are basically embedded by using an embedding of a complete graph even for sparse problem graphs.
This means that the resource of the D-Wave quantum annealer is not exploited efficiently.
In this paper, we propose a fast algorithm for embedding larger subproblems, and we show that better solutions are obtained efficiently by embedding larger subproblems.
\end{abstract}
\begin{document}

\flushbottom
\maketitle
%
%
\thispagestyle{empty}
\section*{Introduction}
Combinatorial optimization problems, the minimization of cost functions with discrete variables, have significant real-world applications.
The cost function of a combinatorial optimization problem can generally be mapped to the Hamiltonian of a classical Ising model \cite{Ising_mapping}.
Inspired by simulated annealing \cite{SA_original}, quantum annealing (QA) \cite{QA_original} was proposed as a method for searching the ground state of a Hamiltonian with a complicated energy landscape.
While SA employs thermal fluctuations to escape local minima, QA utilizes quantum fluctuations.
Numerous studies have investigated whether QA outperforms SA in terms of the computational time required to obtain a high-accuracy solution.
Most of the studies have shown that QA is superior to SA \cite{QA_sup1, QA_sup2, QA_sup3}, while a few have also suggested that it is inferior \cite{QA_inf}.
Recently, a commercial quantum annealer based on superconducting flux qubits \cite{D-wave} has been developed by D-Wave Systems Inc.
Experimental studies using the D-Wave quantum annealer have been performed to compare the performance of QA and SA \cite{D-wave_SA1, QA_sup3} and to demonstrate the applicability of the D-Wave quantum annealer to practical problems \cite{opt_application1, opt_application2, opt_application3}.

The generic form of a time-dependent Hamiltonian in QA is
\begin{equation}
\hat{H} \left( t \right) = A \left( t \right) \hat{H}_{\mathrm{q}} + B \left( t \right) \hat{H}_0, \label{eq:QA_Hamiltonian}
\end{equation}
where $\hat{H}_0$ is the classical Hamiltonian which represents the cost function to be minimized,
and $\hat{H}_{\rm{q}}$ is the quantum fluctuation term for which the ground state is trivial.
At the beginning of QA, the coefficients of the time-dependent Hamiltonian are set to $A(0)=1$, $B(0)=0$, and the system is in trivial ground state determined by $\hat{H}_{\mathrm{q}}$.
At the end of QA, the coefficients are set to $A(\tau)=0$ and $B(\tau)=1$ where $\tau$ is the annealing time.
The system evolves according to the Schr\"{o}dinger equation:
\begin{equation}
i \frac{d}{dt} \ket{\psi(t)} = \hat{H} \left( t \right) \ket{\psi(t)},
\end{equation}
where $\ket{\psi(t)}$ is a state vector of the system and $\hbar$ is set to $1$ for simplicity.
The system will remain closed to the instantaneous ground state of the time-dependent Hamiltonian
if the system changes sufficiently slowly and if the adiabatic condition \cite{adiabatic_condition},
\begin{equation}
\frac{1}{ \left[ \varepsilon_{1} (t) - \varepsilon_{0} (t) \right]^2 } \left| \braket{1(t)| \frac{d \hat{H}(t)}{dt} | 0(t)} \right| \ll 1,
\end{equation}
is always satisfied during QA. Here $\ket{0(t)}, \ket{1(t)}, \varepsilon_{0}(t)$ and $\varepsilon_{1}(t)$ are eigen vectors and eigen energies of the instantaneous ground state and first excited state, respectively.
Thus, by setting the annealing time $\tau$ large enough, we ultimately obtain the ground state of the classical Hamiltonian $\hat{H}_0$, which represents the optimal solution.

\begin{figure}
	\centering
	\includegraphics[width = 0.2\columnwidth]{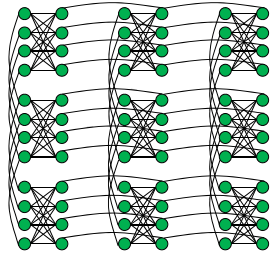}
	\caption{A Chimera graph for $(M, N, L) = (3, 3, 4)$. The nine complete bipartite graphs $K_{4,4}$ are arranged in a grid pattern.}
	\label{fig:chimera}
\end{figure}

The current version of D-Wave quantum annealer (D-Wave 2000Q) implements QA with a transverse magnetic field:
\begin{equation}
\hat{H}_{\mathrm{q}} = - \sum_{i=1}^{N_{\mathrm{q}}} \hat{\sigma}_{i}^{(x)},
\end{equation}
where $N_{\mathrm{q}}$ represents the total number of qubits.
A cost function that the D-Wave quantum annealer can handle is:
\begin{equation}
\hat{H}_{0} = \sum_{(i,j) \in \mathrm{chimera}}^{N_{\mathrm{q}}} J_{ij} \hat{\sigma}_{i}^{(z)} \hat{\sigma}_{j}^{(z)} + \sum_{i=1}^{N_{\mathrm{q}}} h_{i} \hat{\sigma}_{i}^{(z)},
\end{equation}
where the interactions between qubits are restricted to Chimera graph, which is constructed as an $M \times N$ grid of complete bipartite graphs $K_{L,L}$ \cite{Comp_Embed1}.
Chimera graph for $(M, N, L) = (3, 3, 4)$ is shown in Fig. \ref{fig:chimera}, where the vertices and edges represent qubits and the interactions between them, respectively.
Although the Chimera graph for D-Wave 2000Q is $(M, N, L) = (16, 16, 4)$, the number of operable qubits is less than $N_{\mathrm{q}} = 2MNL = 2048$, since there are defects in the qubits and connectivities.

Limited connectivity between the qubits is a restriction to employing the D-Wave quantum annealer for real-world applications.
Before solving an optimization problem, it is necessary to map a problem graph onto a subgraph of the hardware graph.
This process is called \lq\lq minor embedding\rq\rq.
The problem graph is defined as a graph in which the vertices and edges represent the logical variables and interactions between them, respectively.
The hardware graph is defined as a graph for which the vertices and edges represent the qubits and interactions between them, respectively.
It is known to be NP-hard \cite{embed_hard} to find an optimal minor embedding of an arbitrary problem graph into an arbitrary hardware graph.
There exist various algorithms to find the minor embeddings, and a heuristic algorithm proposed by Cai \textit{et al.} \cite{heuristic_Embed} is the most versatile option so far.
While this general algorithm searches for a minor embedding of an arbitrary problem graph into an arbitrary hardware graph, the computational time increases drastically with the number of qubits, especially for sparse problem graphs.
To reduce the computational time for the minor embedding, some algorithms that exploit features of the hardware graphs and specific problem graphs have been developed.
Although the number of logical variables embeddable into hardware graphs is small, utilizing complete graph embedding \cite{Comp_Embed2, Comp_Embed3} is the simplest way to reduce the computation time.
Complete graph embedding can be applied to arbitrary problem graphs with less than 64 logical variables for a Chimera graph with $(M, N, L) = (16, 16, 4)$ without defects.
This is basically the embedding used in {\tt qbsolv} \cite{qbsolv}.
Other embedding algorithms \cite{Virtual_Embed} can find the minor embedding efficiently in reasonably dense problem graphs by exploiting the bipartite structure of the Chimera graph \cite{bipartite_chimera}, and it is possible to embed a larger number of logical variables than for a complete graph embedding \cite{Comp_Embed2, Comp_Embed3}.
A minor embedding of the Cartesian product of two complete graphs, which often appears in real-world optimization problems, has been proposed in the literature \cite{Cartesian_Embed}.
However, efficient embedding algorithms for sparse problem graphs do not exist, despite the fact that sophisticated minor embeddings for sparse problem graphs are more important than for dense problem graphs in order to exploit the potential of the D-Wave quantum annealer.
More logical variables can be embedded with shorter-length chains for sparse problem graphs.

In the methods section below, we propose a fast algorithm for embedding larger subproblems based on Cai's algorithm \cite{heuristic_Embed}.
We do not need to embed all of the logical variables of a problem graph in embedding of subproblems, and the logical variables that can be embedded easily are selected as a part of the subproblem in our proposed algorithm.
As a result, the proposed algorithm can embed larger subproblems than complete graph embedding \cite{Comp_Embed2, Comp_Embed3}, with shorter computational time than the Cai's algorithm \cite{heuristic_Embed}, not only for dense problem graphs but also for sparse problem graphs.

In the following section, we show the improvement in solutions achieved by embedding larger subproblems for a ferromagnetic model and a spin-glass model on a three-dimensional $\pm J$ Ising model with 1,000 logical variables.
Since the cubic lattice with 1,000 variables is not embeddable into D-Wave 2000Q, we extract embeddable subproblems into D-Wave 2000Q and iteratively optimize them using a quantum annealer like {\tt qbsolv}.
In this study, we utilized two algorithms to embed subproblems into the Chimera graph of D-Wave 2000Q, with few defects in the qubits and connectivities.
One is the proposed algorithm, which can embed 380 logical variables, and the other is complete graph embedding \cite{Comp_Embed3}, which can embed only 63 logical variables.
We have confirmed that better solutions can be achieved with less number of iterations for both the ferromagnetic and spin-glass models by embedding large subproblems.

\begin{figure}
	\centering
	\includegraphics[width = 0.6\columnwidth]{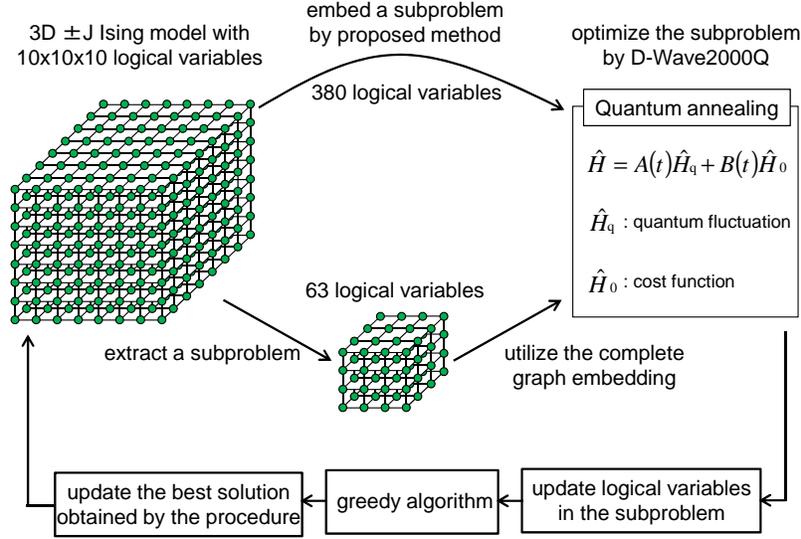}
	\caption{The optimization process implemented in this study. The solutions obtained by utilizing the proposed algorithm and by using complete-graph embedding are compared.}
	\label{fig:evaluation_energy}
\end{figure}
\section*{Results}
In this section, we demonstrate that better solutions are obtained efficiently by embedding larger subproblems for the ferromagnetic and spin-glass models on the three-dimensional $\pm J$ Ising model with $1,000$ logical variables

The optimization process implemented in this study is shown in Fig. \ref{fig:evaluation_energy}.
The problem graph is the cubic lattice with $10 \times 10 \times 10$ logical variables.
We partition the original large problem into subproblems and then iteratively optimize the subproblems using a quantum annealer.
Two algorithms are utilized to embed subproblems in this study.
One is the proposed algorithm, which can embed 380 logical variables, and the other is a complete-graph embedding \cite{Comp_Embed3}, which can embed only 63 logical variables into the Chimera graph of the D-Wave 2000Q with few defects in the qubits and connectivities. 
After embedding and optimizing the subproblem, the logical variables of the subproblem are updated in the output of D-Wave 2000Q.
Then, a greedy algorithm is executed by a conventional digital computer to get to an exact (local) minimum.
In this greedy algorithm, one logical variable is randomly selected, and it is flipped if the energy decreases.
We finish refining the solution using the greedy algorithm if the energy change caused by flipping each logical variable is completely non-negative.
Finally, the best solution obtained by this procedure is updated.
These processes are iterated, and we confirm that the solutions are improved by embedding larger subproblems.

The Hamiltonian optimized in this study is shown below:
\begin{gather}
H_{0} \left( \left\{ x \right\} \right) = \sum_{<i,j>} J_{ij} x_{i} x_{j}, \\
p \left( J_{ij} \right) = p_{\mathrm{F}} \delta \left( J_{ij} - J \right) + \left( 1 - p_{\mathrm{F}} \right) \delta \left( J_{ij} + J \right),
\end{gather}
where $x_{i} \in ( -1, +1 )$ represents a logical variable, $J_{ij}$ is the interaction between nearest neighbors in the cubic lattice, and $p_{\mathrm{F}}$ is the probability that $J_{ij} = +J$, the anti-ferromagnetic interaction.
We evaluated solutions for a ferromagnetic model with $p_{\mathrm{F}}=0.0$ and a spin-glass model with $p_{\mathrm{F}}=0.5$ \cite{3D_phase_annealer}.
The ferromagnetic model has no frustration, so that $x_{1} = x_{2} = \cdots = x_{1000} = -1$ and $+1$ are the trivial ground states.
However, it is often the case that logical variables are divided into two kinds of domains, with the logical variables equal to $+1$ in one domain and $-1$ in the other domain.
The boundaries of the domains are called domain walls, and it is essential to eliminate domain walls to obtain the ground state of ferromagnetic model.
While domain walls cannot be eliminated efficiently by single-spin-flip algorithms such as simulated annealing, cluster Monte Carlo algorithms \cite{cluster_monte_carlo} address domain walls well by flipping logical variables in the same domain simultaneously.
Although the structures of clusters in these algorithms \cite{cluster_monte_carlo} are different from those of the subproblems extracted by the proposed algorithm, we expect that domain walls can be eliminated efficiently by embedding larger subproblems.
In the spin-glass model with $p_{\mathrm{F}}=0.5$, there are many frustrations, and the energy landscape is rugged with many local minima.
In order to obtain better solutions, it is essential to search for as many local minima as possible.
By embedding larger subproblems, the phase space searched by optimizing the subproblem grows exponentially, and it is possible to search for better local minima that could be distant from the current solution in the phase space.
As a result, we expect that better solutions can be obtained efficiently by embedding larger subproblems for both the ferromagnetic and the spin-glass models.

\begin{figure}
	\centering
	\includegraphics[width = 0.8\columnwidth]{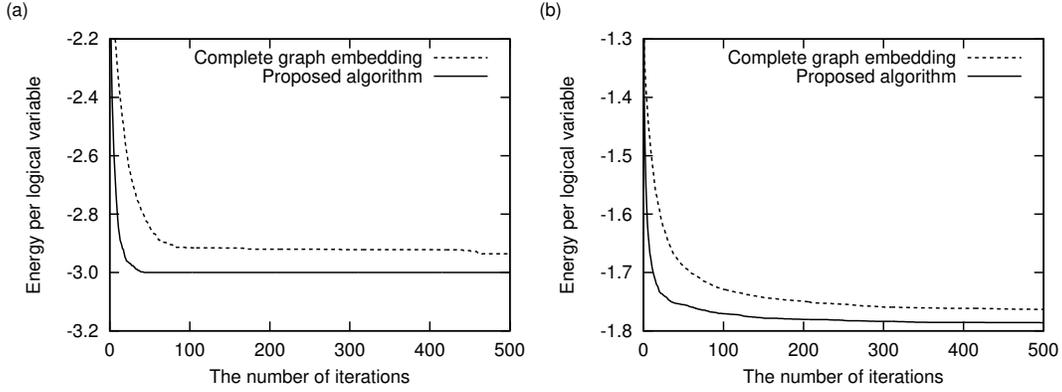}
	\caption{Comparisons of the average energy from 32 trials. (a) The ferromagnetic model with $p_{\mathrm{F}} = 0.0$. (b) The spin-glass model with $p_{\mathrm{F}} = 0.5$.}
	\label{fig:result_energy}
\end{figure}

The energies obtained for $p_{\mathrm{F}} = 0.0$ and $p_{\mathrm{F}} = 0.5$ are shown in Figs. \ref{fig:result_energy}(a) and \ref{fig:result_energy}(b), respectively.
The average energies for 32 trials are plotted, and the same initial states are used for each run [$p_{\mathrm{F}} = 0.0$ and $0.5$, with the two embedding algorithms illustrated in Fig. \ref{fig:evaluation_energy}].
For both $p_{F}=0.0$ and $p_{F}=0.5$, lower energies are obtained with a smaller number of iterations by embedding larger subproblems.
The ground state energy for $p_{F}=0.0$ is $-3$ and the ground state is obtained for all the trials after 45 iterations.

\section*{Discussion}
In the present paper, we showed that better solutions are obtained efficiently by embedding larger subproblems for the spin-glass and ferromagnetic models on the cubic lattice with $10 \times 10 \times 10$ logical variables.
The energy landscape of the spin-glass model is rugged with many local minima.
It is essential to search for as many local minima as possible, and this can be achieved by embedding larger subproblems, for which the phase space is exponentially larger than that of small subproblems.
For the ferromagnetic model, although there are no frustrations and a trivial ground state exists, eliminating the domain walls from which single-spin-flip algorithms suffer is essential to obtain the ground state.
The logical variables in small domains can be flipped simultaneously by embedding larger subproblems, and as a result the ground state can be obtained efficiently with a smaller number of iterations.
Although we demonstrated the improvements in the solutions specifically for the spin-glass and ferromagnetic models on a cubic lattice,
we expect that better solutions can be obtained efficiently for a wide range of optimization problems by embedding larger subproblems.

For practical applications, it is essential to utilize the D-Wave quantum annealer as a part of an iterative solver like {\tt qbsolv}, as long as the problem size embeddable in the D-Wave quantum annealer remains limited.
A hybrid use of classical algorithms and the D-Wave quantum annealer is inevitable for this scheme.
Although we simply adopted a greedy algorithm as a classical optimization algorithm, a myriad of classical algorithms can be combined with the D-Wave quantum annealer \cite{qbsolv, hybrid1, hybrid2}.
One guideline for selecting a classical solver is to exploit the complementary advantages of QA and classical algorithms \cite{qbsolv}.
For example, a more versatile optimization algorithm may be constructed by combining QA and SA \cite{Modernizing}, since QA performs well for the energy landscape with many high and thin barriers, while SA efficiently explores the phase space with low and wide barriers \cite{simple_case}.

Although QA was initially proposed as an optimization method, the D-Wave quantum annealer has recently been considered as a sampling machine.
It has been assumed that the output of D-Wave quantum annealer is close to a Boltzmann distribution of the Hamiltonian at a freeze-out point during annealing \cite{freeze-out},
and applications that utilize the quantum annealer as a sampling machine have been reported \cite{sample1, sample2, sample3, sample4}.
In addition, a local search around a specific initial state using the D-Wave quantum annealer has been proposed in the literature \cite{Modernizing} and it is implemented in D-Wave 2000Q. This is called \lq\lq reverse annealing\rq\rq\cite{reverse}.
By combining reverse annealing and the embedding algorithm proposed in this paper, it may be possible to implement Markov chain Monte Carlo methods efficiently for large problems.

In future work, we will evaluate the utility of embedding larger subproblems for various optimization problems and construct high-performance optimization algorithms that exploit the proposed embedding algorithm.

\section*{Methods}
In this section, we describe a fast algorithm for embedding larger subproblems into a hardware graph.

\subsection*{Definition of minor embedding}
In general, a problem graph is not a subgraph of a Chimera graph, and the problem graph must be mapped onto a subgraph of a Chimera graph in order to solve optimization problems using the D-Wave quantum annealer.
This process is called \lq\lq minor embedding\rq\rq\ of the problem graph into the hardware graph, and this is achieved by representing one logical variable with multiple qubits.
For example, more than two qubits must be assigned to represent a logical variable that interacts with ten logical variables, since the maximum degree of a Chimera graph is six.
If the two qubits $\hat{\sigma}_{1}^{(z)}$ and $\hat{\sigma}_{2}^{(z)}$ are used to represent the same logical variable, $\hat{\sigma}_{1}^{(z)}$ and $\hat{\sigma}_{2}^{(z)}$ must be connected on Chimera graph.
The local energy related to $\hat{\sigma}_{1}^{(z)}$ and $\hat{\sigma}_{2}^{(z)}$ is denoted as $J_{12} \hat{\sigma}_{1}^{(z)} \hat{\sigma}_{2}^{(z)}$, and we can set the local energy of $\hat{\sigma}_{1}^{(z)} = \hat{\sigma}_{2}^{(z)}$ lower than that of $\hat{\sigma}_{1}^{(z)} = - \hat{\sigma}_{2}^{(z)}$ by setting $J_{12} <0$.
If $\left| J_{12} \right|$ is large enough, the optimal solutions of the embedded problem will be identical to that of the original optimization problem.
Note that the assignment of multiple qubits to one logical variable is allowed, while the assignment of multiple logical variables to one qubit is forbidden.
As shown in the literature \cite{Comp_Embed2}, a minor embedding of a problem graph $G_{\mathrm{p}}$ into a hardware graph $G_{\mathrm{q}}$ is defined as follows:
\begin{enumerate}
\item Each vertex $v$ in $V_{\mathrm{p}}$ is mapped to the vertex set of a connected subtree $T_{v}$ of $G_{\mathrm{q}}$.
\item If $(u,v) \in E_{\mathrm{p}}$, then there exist $i_{u}, i_{v} \in V_{\mathrm{q}}$ such that $i_{u} \in T_{u}, i_{v} \in T_{v}$, and $( i_{u}, i_{v} ) \in E_{\mathrm{q}}$.
\end{enumerate}
A connected subtree $T_{v}$ is often called a \lq\lq chain\rq\rq.

\subsection*{A conventional heuristic algorithm}
\begin{figure}
	\centering
	\includegraphics[width = 0.7\columnwidth]{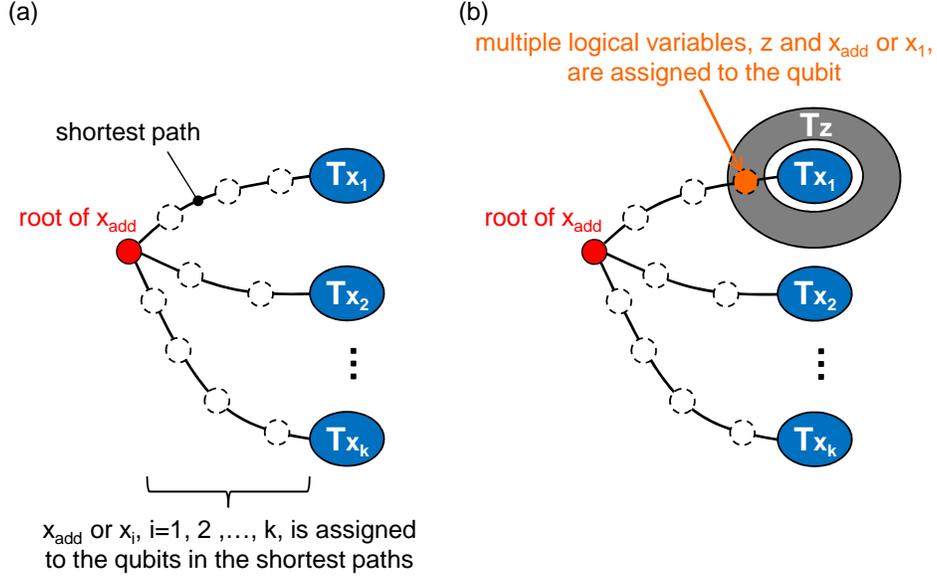}
	\caption{An embedding of a logical variable $x_{\mathrm{add}}$. (a) A case for which paths to the adjacent logical variables already exist. (b) A case for which multiple assignments of logical variables is necessary.}
	\label{fig:heuristic_path}
\end{figure}

A conventional heuristic algorithm for finding a minor embedding of an arbitrary problem graph into an arbitrary hardware graph has been proposed by Cai \textit{et al.} in the literature \cite{heuristic_Embed}.
The embedding process of this algorithm is divided into two stages.
In the initial stage, logical variables are embedded one by one into the hardware graph, and all of the logical variables are embedded by allowing multiple assignments of the logical variables to one qubit.
For example, suppose that the logical variables $x_{1}, ..., x_{k}$ are already embedded in the hardware graph, and a logical variable $x_{\mathrm{add}}$ that is adjacent to $x_{1}, ..., x_{k}$ in the problem graph is selected to be additionally embedded.
In this case, as shown in Fig. \ref{fig:heuristic_path}(a), an unused qubit to which no logical variables are assigned is selected as the root of $x_{\mathrm{add}}$, and the shortest paths from the root of $x_{\mathrm{add}}$ to $T_{x_{1}}, ..., T_{x_{k}}$ are calculated on the hardware graph using Dijkstra's algorithm.
Then, by assigning $x_{\mathrm{add}}$ or $x_{i}$ $(i=1,2,...,k)$ to qubits in the shortest paths, the adjacency between $x_{\mathrm{add}}$ and $x_{1}, ..., x_{k}$ will be represented on the hardware graph.
However, it will often be the case that a path with only unused qubits does not exist.
For example, as shown in Fig. \ref{fig:heuristic_path}(b), if a logical variable $z$ is assigned to all of the qubits adjacent to $T_{x_{1}}$, a path from the root of $x_{\mathrm{add}}$ to $T_{x_{1}}$ with only unused qubits does not exist.
In such a case, by assigning multiple logical variables [$x_{\mathrm{add}}$ or $x_{1}$ and $z$ in Fig. \ref{fig:heuristic_path}(b)] to one qubit, $x_{\mathrm{add}}$ will be embedded once.
After embedding all of the logical variables by allowing multiple assignments in the initial stage, the minor embedding obtained in the initial stage is refined so that only one logical variable is assigned to one qubit in the last stage.

The computational time for this algorithm is dominated by Dijkstra's algorithm.
The computational time $T_{\mathrm{conv}}^{(1)}$ and the number $N_{\mathrm{Dijkstra}}^{(1)}$ of shortest paths searched by Dijkstra's algorithm in the initial stage are given by
\begin{gather}
T_{\mathrm{conv}}^{(1)} \sim O \left( N_{\mathrm{Dijkstra}}^{(1)} T_{\mathrm{Dijkstra}} \right), \label{eq:T1_conv} \\
N_{\mathrm{Dijkstra}}^{(1)} \sim O \left( | E_{\mathrm{p}} | \right),
\end{gather}
and $T_{\mathrm{conv}}^{(2)}$ and $N_{\mathrm{Dijkstra}}^{(2)}$ in the last stage are given by
\begin{gather}
T_{\mathrm{conv}}^{(2)} \sim O \left( N_{\mathrm{Dijkstra}}^{(2)} T_{\mathrm{Dijkstra}} \right), \\
N_{\mathrm{Dijkstra}}^{(2)} \sim O \left( | V_{\mathrm{p}} | | V_{\mathrm{q}} | | E_{\mathrm{p}} | \right),\end{gather}
where $T_{\mathrm{Dijkstra}}$ represents the computational time for Dijkstra's algorithm:
\begin{equation}
T_{\mathrm{Dijkstra}} \sim O \left( | E_{\mathrm{q}} | + | V_{\mathrm{q}} | \log | V_{\mathrm{q}} | \right).
\end{equation}
Here, $|V_{\mathrm{p}}|$ and $|E_{\mathrm{p}}|$ are the number of vertices and edges in the problem graph, and $|V_{\mathrm{q}}|$ and $|E_{\mathrm{q}}|$ are the number of vertices and edges in the hardware graph, respectively.
In this algorithm, the vertex-weighted shortest paths are searched in order to distinguish used and unused qubits.
The computational time in the last stage is obviously dominant.
So we expect that the computational time will be drastically reduced by avoiding multiple assignments of logical variables in the initial stage, since the implementation of the last stage becomes unnecessary.

\subsection*{Proposed algorithm}
Here we focus on the embedding of subproblems and propose a fast algorithm to find minor embeddings of subproblems.
In embedding a subproblem, we can select logical variables that are embeddable without multiple assignments as a part of the subproblem, since it is not necessary to embed all of the logical variables included in the problem graph.
While in a conventional algorithm, the search for a minor embedding is subject to a strong restriction, that all of the logical variables included in the problem graph must be embedded.
The multiple assignments of logical variables are mainly caused by this restriction.
\begin{figure}
	\centering
	\includegraphics[width = 0.9\columnwidth]{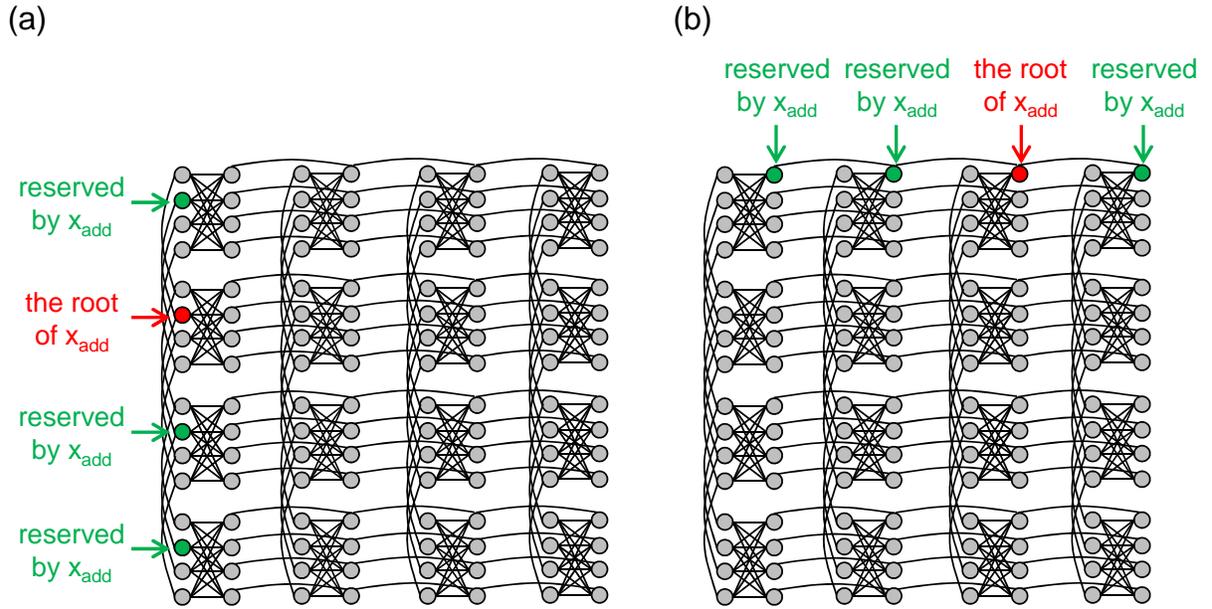}
	\caption{ Examples of reserved qubits associated to the root of $x_{\mathrm{add}}$. (a) An example showing vertically reserved qubits associated to the root of $x_{\mathrm{add}}$. (b) An example showing horizontally reserved qubits associated to the root of $x_{\mathrm{add}}$.}
	\label{fig:reserve}
\end{figure}
\begin{figure}
 \centering
 \includegraphics[width = 0.7\columnwidth]{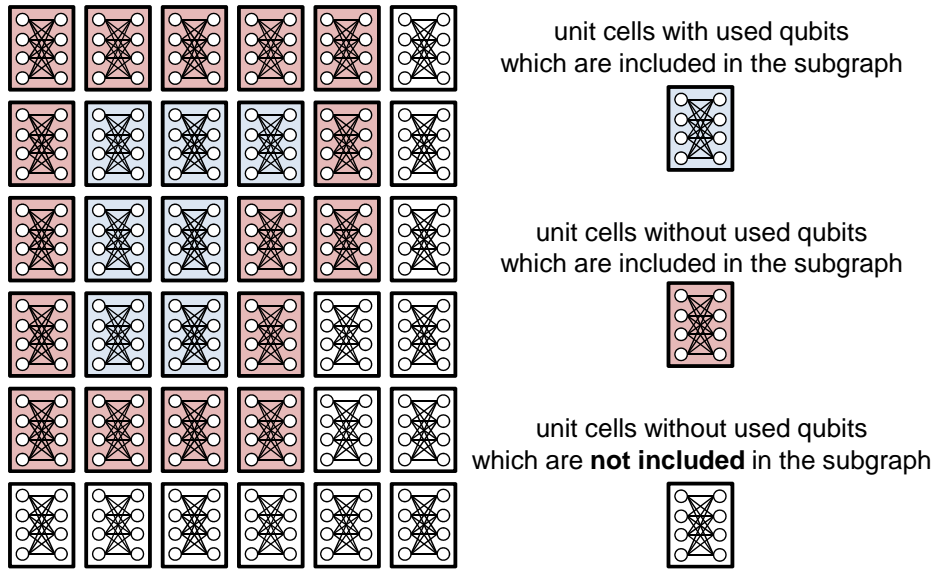}
 \caption{An example of the unit cells included in a subgraph. The unit cells colored blue and red are included. The number of qubits in the subgraph is dominated by the qubits in the unit cells colored red.}
 \label{fig:prop_subgraph}
\end{figure}

However, for dense problem graphs, the logical variables embeddable without multiple assignments become extinct before all the qubits are exploited.
To mitigate this issue, the proposed algorithm includes a reservation system that leaves space to extend the chains.
As shown in Fig. \ref{fig:reserve}(a), if a qubit on the left side of a complete bipartite graph $K_{4,4}$ in Chimera graph is selected as the root of $x_{\mathrm{add}}$, qubits to extend the chain vertically are reserved by $x_{\mathrm{add}}$, and assignment of other logical variables to these qubits are forbidden.
If a qubit on the right side is selected as the root of $x_{\mathrm{add}}$, qubits to extend the chain horizontally are reserved by $x_{\mathrm{add}}$, as shown in Fig. \ref{fig:reserve}(b).
The reserved qubits are released after the embedding of all the logical variables adjacent to $x_{\mathrm{add}}$ are completed.

The refinement of the embedding in the last stage of the conventional algorithm can be eliminated in the proposed algorithm.
In addition, the breadth-first search in a subgraph of a hardware graph consisting only of unused qubits is sufficient to search the shortest paths, since multiple assignments are forbidden.
The computational time $T_{\mathrm{prop}}$ for the proposed algorithm and the number $N_{\mathrm{breadth}}$ of the shortest paths searched by the breadth-first search are given by
\begin{gather}
T_{\mathrm{prop}} \sim O \left( N_{\mathrm{breadth}} T_{\mathrm{breadth}} \right), \\
N_{\mathrm{breadth}} \sim | E_{\mathrm{p}} |.
\end{gather}
The computational time $T_{\mathrm{breadth}}$ for the breadth-first search is given by
\begin{equation}
T_{\mathrm{breadth}} \sim O \left( | e_{\mathrm{q}} | \right),
\end{equation}
where $| e_{\mathrm{q}} |$ is the number of edges included in the subgraph of the hardware graph with unused qubits.
For Chimera graph, as shown in Fig. \ref{fig:prop_subgraph}, it is sufficient to consider unit cells with used qubits and adjacent unit cells without used qubits.
The maximum number of edges included in the subgraph is limited to
\begin{equation}
|e_{\mathrm{q}}| \sim O( |V_{\mathrm{q}}|^{1/2} L^{3/2} ).
\end{equation}

The embedding algorithm proposed in this section does not strongly depend on the structure of the hardware graphs, except for the reservation system, and we can easily adapt the embedding algorithm for other hardware graphs.

\begin{figure}
 \centering
 \includegraphics[width = 0.8\columnwidth]{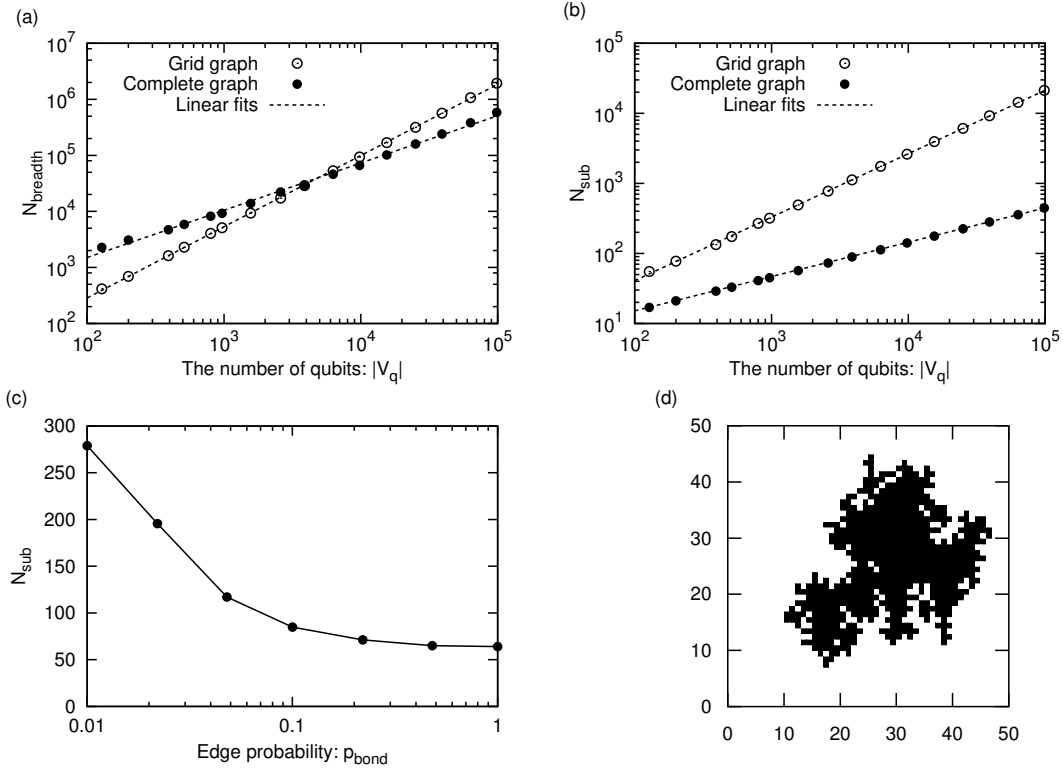}
 \caption{Experimental results from the proposed algorithm. (a) The $|V_{\mathrm{q}}|$ dependence of $N_{\mathrm{breadth}}$. (b) The $|V_{\mathrm{q}}|$ dependence of $N_{\mathrm{sub}}$. (c) The $p_{\mathrm{bond}}$ dependence of $N_{\mathrm{sub}}$. (d) An example of a subproblem embedded into D-Wave 2000Q.}
 \label{fig:result_embed}
\end{figure}
\subsection*{Experimental results}
In order to confirm the scalability of our algorithm, we evaluated the $|V_{\mathrm{q}}|$ dependence of the number $N_{\mathrm{breadth}}$ of the shortest paths searched by the breadth-first search and the size $N_{\mathrm{sub}}$ of the embedded subproblems.
We have used the proposed algorithm to embed subproblems of a grid graph with $300 \times 300$ variables and a complete graph with $1,000$ variables into a Chimera graph with $10^{2} \sim 10^{5}$ qubits.
The results for $N_{\mathrm{breadth}}$ are shown in Fig. \ref{fig:result_embed}(a).
Linear fits to the experimental results yield
\begin{gather}
N_{\mathrm{breadth}}^{\mathrm{(grid)}} \sim O \left( |V_{\mathrm{q}}|^{1.27} \right), \\
N_{\mathrm{breadth}}^{\mathrm{(complete)}} \sim O \left( |V_{\mathrm{q}}|^{0.84} \right).
\end{gather}
The $|V_{\mathrm{q}}|$ dependence of $N_{\mathrm{breadth}}$ is less than $O \left( |V_{\mathrm{q}}|^{1.3} \right)$, even for a grid graph with sparse connectivity.
As the exponent is not large, we expect the proposed algorithm to be feasible even if the number of qubits is increased in a future version of the D-Wave quantum annealer.
The results for $N_{\mathrm{sub}}$ are shown in Fig. \ref{fig:result_embed}(b).
Linear fits to the experimental results yield
\begin{gather}
N_{\mathrm{sub}}^{\mathrm{(grid)}} \sim O \left( |V_{\mathrm{q}}|^{0.91} \right), \\
N_{\mathrm{sub}}^{\mathrm{(complete)}} \sim O \left( |V_{\mathrm{q}}|^{0.50} \right).
\end{gather}
The size of subproblems for the complete graph $N_{\mathrm{sub}}^{\mathrm{(complete)}}$ is identical to the maximum size embeddable into a Chimera graph.
Although the sizes $N_{\mathrm{sub}}^{\mathrm{(grid)}}$ of the subproblems for the grid graph are smaller than the ideal dependence $O \left( |V_{\mathrm{q}}|^{1.0} \right) $,
they are much larger than those for the complete graph.
These results imply that subproblems larger than the complete graph embedding \cite{Comp_Embed2, Comp_Embed3} can be embedded depending on the connectivity of the problem graphs, and with a computation time shorter than that required for Cai's algorithm \cite{heuristic_Embed}.
Because the subproblem embedding is searched greedily and refinement of the embedding is not implemented, an optimal embedding is hardly found especially for sparse problem graphs.
However, the computational time is drastically reduced.

The sizes of subproblems extracted from an Erd\H{o}s-R\'{e}nyi model with $1,000$ logical variables for various edge probabilities $p_{\mathrm{bond}}$ are shown in Fig. \ref{fig:result_embed}(c).
Interactions between variables are randomly generated in this model, with edge probability $p_{\mathrm{bond}}$, and the average sizes of subproblems for $100$ instances are plotted in the graph.
Subproblems are embedded into a Chimera graph in D-Wave 2000Q.
As $p_{\mathrm{bond}}$ decreases, the size of the embedded subproblem increases.
The proposed algorithm can embed larger subproblems depending on the connectivity of the problem graphs even if the interactions between variables are randomly generated.

An example of a subproblem extracted from a grid graph with $50 \times 50$ logical variables is shown in Fig. \ref{fig:result_embed}(d).
The subproblem is embedded into a Chimera graph of the D-Wave 2000Q.
The logical variables embedded as the subproblem are colored black.
It is desirable that an extracted subproblem consists not of tree structures that are easily optimized but instead of many closed loops that can contain frustrations.
The subproblem shown in Fig. \ref{fig:result_embed}(d) satisfies this condition.

\bibliography{sample}

\section*{Acknowledgements}
The authors are deeply grateful to Shu Tanaka, Masamichi J. Miyama, Tadashi Kadowaki, Hirotaka Irie and Akira Miki for fruitful discussions.
One of the authors M. O. is grateful to the financial support from JSPS KAKENHI 15H03699 and 16H04382, the JST-START, JST-CREST(No.JPMJCR1402), and the ImPACT program.

\section*{Author contributions statement}
S. O. conceived and developed the concept, and carried out all the experiments. M. O. proposed the plan to evaluate the validity of the concept, discussed the details of the results and reviewed the manuscript.
M. T. and S. T. directed the project in our study.

\section*{Additional information}
\textbf{Competing interests:} The authors declare no competing interests.
\end{document}